\documentclass[sigconf]{acmart}
\AtBeginDocument{%
  }

\setcopyright{acmcopyright}
\copyrightyear{2025}
\acmYear{2025}
\acmISBN{978-1-4503-XXXX-X/2025/07}

\begin{document}

\title{DLiPath: A Benchmark for the Comprehensive Assessment of Donor Liver Based on Histopathological Image Dataset}

\author{Liangrui Pan}
\affiliation{%
  \institution{Hunan University}
  \city{Changsha}
  \state{Hunan}
  \country{China}
}
\email{panlr@hnu.edu.cn}

\author{Xingchen Li}
\affiliation{%
  \institution{Hunan University}
  \city{Changsha}
  \country{Hunan}}
\email{lxc\_stars@gs.zzu.edu.cn}

\author{Zhongyi Chen}
\affiliation{%
	\institution{The Third Xiangya Hospital of Central South University}
	\city{Changsha}
	\country{Hunan}}
\email{15585000531@163.com}

\author{Ling Chu}
\authornote{Corresponding author.}
\affiliation{%
  \institution{The Third Xiangya Hospital of Central South University}
  \city{Changsha}
  \country{Hunan}}
\email{chu156@csu.edu.cn}

\author{Shaoliang Peng}
\affiliation{%
  \institution{Hunan University}
  \city{Changsha}
  \country{Hunan}}
\email{slpeng@hnu.edu.cn}
\authornotemark[1]

\renewcommand{\shortauthors}{Liangrui Pan et al.}

\begin{abstract}
  Pathologists’ comprehensive evaluation of donor liver biopsies provides crucial information for accepting or discarding potential grafts. However, rapidly and accurately obtaining these assessments intraoperatively poses a significant challenge for pathologists. Features in donor liver biopsies, such as portal tract fibrosis, total steatosis, macrovesicular steatosis, and hepatocellular ballooning are correlated with transplant outcomes, yet quantifying these indicators suffers from substantial inter‑ and intra‑observer variability. To address this, we introduce DLiPath, the first benchmark for comprehensive donor liver assessment based on a histopathology image dataset. We collected and publicly released 636 whole slide images from 304 donor liver patients at the Department of Pathology, the Third Xiangya Hospital, with expert annotations for key pathological features (including cholestasis, portal tract fibrosis, portal inflammation, total steatosis, macrovesicular steatosis, and hepatocellular ballooning). We selected nine state‑of‑the‑art multiple‑instance learning (MIL) models based on the DLiPath dataset as baselines for extensive comparative analysis. The experimental results demonstrate that several MIL models achieve high accuracy across donor liver assessment indicators on DLiPath, charting a clear course for future automated and intelligent donor liver assessment research. Data and code are available at \url{https://github.com/panliangrui/ACM_MM_2025}.
  
\end{abstract}


\begin{teaserfigure}
  \includegraphics[scale=0.63]{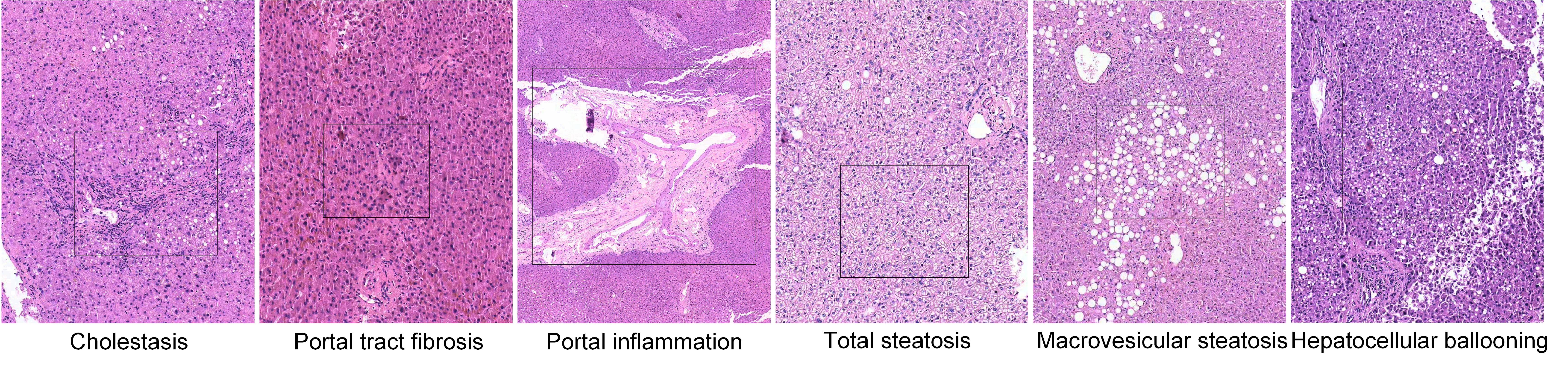}
  \caption{Six pathological indicators for donor liver assessment.}
  \label{fig:teaser}
\end{teaserfigure}


\maketitle

\section{Introduction}
As the most effective treatment for end‑stage liver disease, liver transplantation’s success is highly dependent on precise assessment of donor liver quality \cite{qincell}, \cite{zhang2024cell}. With the global incidence of metabolic diseases on the rise, the proportion of steatotic donor livers has increased significantly to approximately 30\% of deceased grafts \cite{ahmed2021liver}, \cite{gedallovich2022liver}. The demonstrated correlation between these grafts and higher rates of primary non‑function (PNF), ischemia‑reperfusion injury (IRI), and recipient mortality highlights the urgent need to develop a systematic framework for donor liver evaluation. To date, donor evaluation has evolved from purely morphological examination to an integrated analysis of multidimensional pathophysiological indicators \cite{agopian2018evaluation}, \cite{xiang2024current}. Among these, cholestasis, portal tract fibrosis, portal inflammation, total steatosis, macrovesicular steatosis, and hepatocellular ballooning are closely correlated with post‑transplant complications and have become core parameters for optimizing graft selection \cite{gedallovich2022liver}.

Steatotic degeneration, particularly macrovesicular steatosis exceeding 30\%, has been shown to exacerbate cold ischemia–reperfusion injury through mechanisms such as impaired sinusoidal microcirculation and mitochondrial dysfunction \cite{chu2016steatotic}, \cite{chu2013impact}. Studies indicate that donor livers with moderate macrovesicular steatosis carry a threefold higher risk of primary non‑function compared to non‑steatotic grafts, while severe steatosis (>60\%) is often considered a contraindication to transplantation \cite{sharkey2011high}, \cite{nocito2006steatosis}. However, reliance solely on the extent of steatosis has limitations. For example, portal tract fibrosis can independently cause graft hypoperfusion by promoting sinusoidal narrowing and collagen deposition, and portal inflammation intensifies oxidative stress after ischemia reperfusion injury through activation of the TLR4 signaling pathway\cite{arslan2010tlr2}. In addition, cholestasis, which is often linked to biliary anatomical variations and prolonged cold ischemia time, may lead to postoperative biliary strictures or bile leakage.

Deep learning (DL) demonstrates powerful learning capacity and excels in handling complex pathological features. Consequently, DL has been widely applied in computational pathology \cite{madabhushi2016image}, \cite{cui2021artificial}. Current research focuses on developing fully automated donor liver assessment methods based on DL algorithms for quantitative histological analysis \cite{takahashi2023artificial}, \cite{sun2020deep}, \cite{d2020histological}, \cite{bhat2023artificial}. DL approaches can reduce both inter‑ and intra‑observer variability in cellular-structure assessment and improve result reproducibility. Leveraging these advantages, numerous DL‑based methods for donor liver assessment have been proposed to advance this field. For example, Sun et al. developed a convolutional neural network (CNN) that generates steatosis probability maps from whole‑slide images (WSIs) of hematoxylin-eosin stained (H\&E) frozen sections and then computes the percentage of steatosis \cite{sun2020deep}. A comparison between pathologists’ scores and those produced by the CNN specifically designed for automated steatosis assessment showed consistency in intraclass correlation coefficient (ICC) measurements, though this did not reach statistical significance \cite{gambella2024improved}. Additionally, a DL model based on radiomics and multiple networks has been used for hepatocellular‑carcinoma liver‑transplantation risk assessment \cite{he2021imageomics}. These methods enable pathologists to perform consistent and reproducible histopathological assessments, reducing the burden of time‑consuming and repetitive tasks. In summary, to date there is still no fully reliable automated strategy for donor liver assessment.

To address this urgent clinical need, we collected 636 WSIs from 304 donor liver patients at the Department of Pathology of the Third Xiangya Hospital of Central South University, constructed DLiPath dataset, and expert annotations were performed for key pathological features such as cholestasis, portal tract fibrosis, portal inflammation, total steatosis, macrovesicular steatosis, and hepatocellular ballooning. Our goal is to enable a comprehensive assessment of donor liver suitability using WSIs. Based on DLiPath, we established six assessment indicators whose grades are closely correlated with graft usability. We then formulated the assessment task as a multi-classification problem under a multiple‐instance learning (MIL) framework. Importantly, this study employs nine MIL‐based models on DLiPath to systematically assess these donor liver criteria in histopathological images and discusses the potential of DL models for precision medicine. MIL‑based donor liver assessment can reduce pathologists’ subjective bias while improving the accuracy and timeliness of graft assessment, thereby advancing the field of computational pathology. In summary, our contributions are as follows: 

1) \textbf{Donor Liver Histopathology Image Dataset:}  
	DLiPath consists of 636 WSIs of 304 patients who underwent allogeneic liver transplantation at Department of Pathology, the Third Xiangya Hospital of Central South University, and it provides both the original WSIs (available upon request) and image feature representations derived from a pretrained model.
	
2) \textbf{Annotation of Assessment Criteria:}  
	Multiple pathologists annotated every WSI in the DLiPath dataset, labeling key features relevant to donor liver assessment, including cholestasis, portal tract fibrosis, portal inflammation, total steatosis, macrovesicular steatosis, and hepatocellular ballooning.
	
3) \textbf{Comprehensive Benchmarking:}  
	We performed extensive benchmarking of six donor liver assessment tasks on the DLiPath dataset using nine MIL methods, demonstrating the potential of MIL approaches for accurate and objective assessment of donor liver histopathology images.
	

\section{Related work}

\subsection{Liver Histopathology Image Dataset}
Although several public datasets have emerged in the field of liver histopathology image analysis to meet various needs, including classification, segmentation and multi‑omics association studies, most of them focus on hepatocellular carcinoma rather than donor livers. For example, The Cancer Genome Atlas Liver Hepatocellular Carcinoma integrates 491 WSIs of hepatocellular carcinoma (HCC) from multiple centers, along with matched clinical and genomic data, supporting tumor-microenvironment and phenotype–genotype correlation analyses \cite{damrauer2021genomic}. The Hepatic Histopathology Dataset provides hundreds of H\&E-stained WSIs and high-resolution patches, all manually annotated with cellular- and tissue-level features, and is often used as a benchmark for segmentation and classification \cite{lekshmi2020hepatic}.  The PAIP 2019 Liver Cancer Segmentation Challenge offers 40 WSIs with two-level expert annotations of tumor lesions and viable tumor regions, serving as a benchmark for tumor segmentation and viability estimation \cite{kim2021paip}. A dataset of 522 histopathological images, annotated with vascular structures, microvascular invasion (MVI) regions, and HCC grades, has driven research in pathological vessel segmentation and automatic MVI detection. Additionally, the HCC Dataset from Coimbra University Hospital (Kaggle/UCI) includes clinical slide images and survival data for 165 patients, providing real clinical samples for survival prediction and classification researches \cite{santos2015new}. The above datasets range from multi-center large-scale WSIs to specific pathology annotation and task-specific analysis, fully supporting the development and evaluation of various algorithms in the field of liver tissue pathology image analysis. However, they concentrate almost exclusively on liver cancer, leaving a gap in datasets for donor liver assessment. To fill this void, we collaborated with pathologists and proposed a donor liver assessment dataset to promote research and progress in this direction.
\subsection{Multiple-instance learning in digital pathology}
The processing of histopathology images has evolved through three main stages: traditional image processing techniques, machine learning algorithms, and DL with weak supervision. Today, the predominant approach is MIL, a weakly supervised framework in which training samples are organized as “bags”, each containing multiple unlabeled instances, and only the bag as a whole is assigned a label. MIL methods can be divided into two categories: one that obtains the final bag‐level prediction directly from instance‐level predictions, and another that achieves bag‐level prediction by aggregating instance features \cite{cheplygina2015label,coudray2018classification,li2021dual,wang2019weakly,hou2016patch,lin2022interventional}. In the instance‐based category, each instance in the bag is scored or classified individually, and then operations such as max‐pooling or mean‐pooling are used to combine the instance‐level predictions into a bag‐level label \cite{shao2021transmil,qian2022transformer,yao2020whole,zhang2022dtfd,zhao2020predicting}. These methods require an instance‐level classifier, but in the absence of true instance labels they often suffer from noise accumulation. For example, ABMIL employs a learnable attention mechanism that assigns weights to instances before computing a weighted sum, significantly improving weakly supervised WSI classification performance \cite{ilse2018attention}. DSMIL is a dual‐stream multiple‐instance learning network that first uses max‐pooling to identify key instances and then applies a self‐attention stream to aggregate their features, with performance further enhanced by self‐supervised contrastive learning \cite{li2021dual}. CLAM builds on attention‐based aggregation by adding an instance‐level clustering constraint, learning class‐specific attention and refining the feature space via clustering \cite{lu2021data}. TransMIL applies the Transformer architecture directly to the sequence of instances, encoding relationships through self‐attention and aggregating them into a bag representation \cite{shao2021transmil}. Recently, PSMIL aligned instance distributions in probability space to address feature‐space drift, offering a new aggregation paradigm for deep MIL \cite{durethinking}. However, given the high dimensionality and abundance of features in histopathology images, MIL methods may benefit from incorporating multi‐scale features to further boost downstream task accuracy.

\section{DLiPath Dataset}
\subsection{Dataset Collection}

In the period from January 2020 to October 2024, this study at the Third Xiangya Hospital enrolled 304 adult patients who underwent allogeneic liver transplantation. For each patient, one frozen section and several paraffin‐embedded sections were obtained, along with the corresponding clinical data. In total, 636 histopathology images were collected and scanned. Three pathology experts reviewed a subset of each patient’s WSIs and scored the donor livers for chronic hepatitis using the Scheuer grading system \cite{scheuer1991classification}. Inclusion criteria were: (1) adult recipients who underwent allogeneic liver transplantation at our hospital between January 2020 and October 2024; (2) donor liver tissue that had completed standard pathological examination prior to transplantation, with intact paraffin-embedded blocks or high-quality frozen samples available; (3) successful digitization of histological slides without significant staining dropout, folding, or scanning artifacts; and (4) complete perioperative clinical records, including donor–recipient matching data, surgical records, and postoperative follow-up information. Exclusion criteria were: (1) insufficient donor liver sample volume or inability to produce qualified digital slides; (2) missing key clinical information (e.g., unrecorded cold ischemia time of donor liver or unknown postoperative survival status of recipient); (3) presence of artifacts in digital slides that interfere with assessment (e.g., bubbles, knife marks, or over-staining obscuring critical structures); (4) combined multi-organ transplantation or recipients undergoing re-transplantation; and (5) explicit refusal by the patient or family to use pathological data for research purposes.

\subsection{Dataset Organization}
The main indicators of donor liver histopathological asseassessment include cholestasis, portal tract fibrosis, portal inflammation, total steatosis, macrovesicular steatosis, and hepatocellular ballooning. Multiple pathologists will score the histopathology images based on their years of diagnostic experience; if two junior pathologists agree on a score, that label is assigned. If their scores differ, a senior pathologist is consulted to make the final determination. Each indicator is graded on a multiple point scale: none, mild, moderate, and severe. The scoring criteria for these indicators are listed in Table~\ref{tab:labels}.

\begin{table}[htbp]
	\centering
	\caption{Histopathological Asseassessment Indicators for Post‐mortem Donor Liver Biopsy.}
	\label{tab:labels}
	\scalebox{0.7}{\begin{tabular}{lcccc}
		\hline
		\textbf{Indicator} & \textbf{None} & \textbf{Mild} & \textbf{Moderate} & \textbf{Severe} \\
		\hline
		Cholestasis               & None             & Mild (5\%–30\%)     & Moderate (30\%–50\%)  & Severe (>50\%)       \\
		Portal tract fibrosis     & None             & Mild                & Moderate              & Severe               \\
		Portal inflammation       & None             & Mild                & Moderate              & Severe               \\
		Overall steatosis         & None             & Mild (5\%–30\%)     & Moderate (30\%–60\%)  & Severe (>60\%)       \\
		Macrovesicular steatosis  & None             & Mild (5\%–30\%)     & Moderate (30\%–60\%)  & Severe (>60\%)       \\
		Hepatocellular ballooning & None             & Mild (5\%–30\%)     & Moderate (30\%–60\%)  & Severe (>60\%)       \\
		\hline
	\end{tabular}
}
\end{table}

\subsection{Data Preprocessing}
The histopathological images were saved as WSI in svs format based on a 20$\times$ scanning magnification. First, we used the Otsu algorithm to distinguish the tissue regions of the WSI and remove the background regions. Then, we used the sliding window strategy to divide the WSI into $256 \times 256$ pixel patches based on the field of view under a 20$\times$ microscope and marked the spatial coordinates of the patches \cite{joo2023classification}. In order to extract the features of the patches, we used CTransPath to extract one-dimensional features for each patch to enrich the diverse feature representation of WSI \cite{wang2022transformer}. The histopathological image pre-training model based on the Swin Transformer architecture of CTransPath is mainly used for patch feature extraction \cite{liu2021swin}. Among them, the convolution module and the transformer module can extract local features in the image data and capture global features and long range dependencies \cite{pan2024feature}. Based on CTransPath, the embedded features of the patches are extracted, and each patch embedding is unified into a 768-dimensional vector representation. Therefore, the feature vector form of all patches in the WSI can be represented as $V = (m,768)$. $m$ represents the number of patches in the WSI.

\section{Experiments and Benchmarks}
\subsection{Task Definition}
We define the six indicators of donor liver histopathology assessment as a multi-classification task, that is, learning a model on a given dataset $D = \{ ({x_i},{y_i})\} _{i = 1}^N$ so that each sample ${x_i} \in {R^d}$ is correctly mapped to a category ${y_i}$ in the label space $y = \{ 1,...,n\} $. The label is usually represented as a vector ${y_i} \in {\{ 1,n\} ^n}$ in one-hot encoding, where only the ${y_i}$-th component is 1 and the rest are 0. For the multi-classification task, the MIL model can be represented as a mapping function ${f_\theta }:{R^d} \to {R^n}$, whose output $z = {f_\theta }(x)$ is called "logits". The logits are converted to a predicted probability distribution through the Softmax function:
${p_j}(x;\theta ) = \frac{{\exp ({z_j})}}{{\sum\nolimits_{k = 1}^n {\exp ({z_k})} }},j = 1,...,n,$
Where, $\sum\nolimits_{j = 1}^n {{p_i}}  = 1$ and ${p_i} \ge 0$.

\subsection{Baselines Methods}
To comprehensively evaluate the DLiPath dataset and assess the potential of donor liver histopathology images, we employed nine state-of-the-art (SOTA) MIL models to predict six key indicators within DLiPath. These MIL models are ABMIL \cite{ilse2018attention}, CLAM-SB \cite{lu2021data}, CLAM-MB \cite{lu2021data}, DSMIL \cite{li2021dual}, ACMIL \cite{zhang2024attention}, DGRMIL \cite{zhu2024dgr}, IBMIL \cite{lin2023interventional}, ILRA \cite{xiang2023exploring}, and TransMIL \cite{shao2021transmil}, all of which have demonstrated strong analytic performance on segmentation and classification tasks across multiple histopathology image datasets. Please refer to the supplementary material for detailed model definitions.

\subsection{Evaluation Metrics}
In our experiments, we used AUC, Accuracy, Precision, Recall, and F1-Score as evaluation metrics for the model’s predictions of cholestasis, Portal tract fibrosis, portal inflammation, total steatosis, macrovesicular steatosis, and hepatocellular ballooning. Higher values of AUC, Accuracy, Precision, Recall, and F1-Score indicate better model performance. All evaluation metrics are reported as the mean over five-fold cross-validation to evaluate model robustness.

\section{Results}
\subsection{Assessment of Cholestasis Based on MIL Models}

Table~\ref{tab:bile_stasis} summarizes the comparative performance of nine MIL-based architectures on the cholestasis classification task. Among these, TransMIL achieves the highest accuracy of 0.9423 and AUC of 0.9781, underscoring the strength of its Transformer‑based correlated MIL framework in modeling both spatial and semantic dependencies across instances. Its F1‑Score of 0.9444 further reflects a well‑balanced trade‑off between sensitivity and specificity. The next best models, DGRMIL and DSMIL, attain AUCs of 0.9232 and 0.8943 with accuracies of 0.8125 and 0.7684, respectively. DGRMIL’s global‑vector cross‑attention mechanism yields a high recall of 0.9411 and an F1‑Score of 0.8566, indicating robust positive–negative discrimination, while DSMIL’s dual‑stream maximal self‑attention effectively identifies salient instances under limited-data conditions. Mid‑tier MIL methods such as ACMIL, CLAM-SB, and CLAM-MB also show fairly high evaluation accuracy. In contrast, IBMIL and ILRA underperform, with accuracies of 0.5329 and 0.4989 and AUCs of 0.6189 and below the random‑guess baseline, respectively. These results suggest that the causal‑intervention strategy in IBMIL and the iterative low‑rank attention in ILRA require further refinement to capture discriminative features in a multiple class setting. Overall, TransMIL and DGRMIL emerge as the top performers, while CLAM variants and DSMIL provide balanced solutions for scenarios prioritizing interpretability or data efficiency, and IBMIL/ILRA highlight promising avenues for future enhancement.

\begin{table}[ht]
	\centering
	\caption{Performance comparison of different MIL models on cholestasis assessment.}
	\label{tab:bile_stasis}
		\scalebox{0.85}{\begin{tabular}{lccccc}
		\toprule
		Models       & Accuracy & AUC    & Precision & Recall  & F1-Score \\
		\midrule
		ABMIL \cite{ilse2018attention}       & 0.8864   & 0.4510 & 0.9022    & 0.9545  & 0.9273   \\
		CLAM-SB \cite{lu2021data}     & 0.6532   & 0.7215 & 0.6843    & 0.7728  & 0.7252   \\
		CLAM-MB \cite{lu2021data}     & 0.5852   & 0.6739 & 0.7138    & 0.7568  & 0.7115   \\
		DSMIL \cite{li2021dual}       & 0.7684   & 0.8943 & 0.7211    & 0.9355  & 0.8142   \\
		ACMIL \cite{zhang2024attention}       & 0.7255   & 0.6825 & 0.8015    & 0.6537  & 0.7193   \\
		DGRMIL \cite{zhu2024dgr}      & 0.8125   & 0.9232 & 0.7855    & 0.9411  & 0.8566   \\
		IBMIL \cite{lin2023interventional}       & 0.5329   & 0.6189 & 0.5874    & 0.6642  & 0.6237   \\
		ILRA \cite{xiang2023exploring}        & 0.4989   & 0.5822 & 0.5439    & 0.6215  & 0.5797   \\
		TransMIL \cite{shao2021transmil}    & 0.9423   & 0.9781 & 0.9254    & 0.9636  & 0.9444   \\
		\bottomrule
	\end{tabular}
}

\end{table}

\subsection{Assessment of Portal Tract Fibrosis Based on the MIL Models}

To evaluate the performance of MIL models in assessing portal tract fibrosis, we examine each model’s results on five metrics as shown in Table~\ref{tab:portal fibrosis}. Overall, TransMIL and ACMIL achieve the best performance. Specifically, TransMIL attains the highest Accuracy of 0.9318 and AUC of 0.9564, indicating that its Transformer architecture’s ability to capture spatial and semantic correlations among instances substantially enhances weakly supervised slide-level classification. Furthermore, its Precision, Recall, and F1-Score of 0.9425, 0.9282, and 0.9356, respectively, demonstrate high robustness in identifying both positive and negative samples. Close behind, ACMIL achieves an Accuracy of 0.9125 and an AUC of 0.8841, showing its effectiveness in preventing attention collapse and improving generalization. Its Precision of 0.9316 and Recall of 0.8964 indicate that ACMIL can increase key-instance coverage while maintaining a low false-positive rate, and its F1-Score of 0.9138 further confirms the balanced advantage of this strategy for subtype classification. DGRMIL and DSMIL also perform well, with Accuracies of 0.8955 and 0.9113 and AUCs of 0.8577 and 0.8238, respectively, illustrating that global-vector diversification and dual-stream maximal self-attention architectures can also effectively integrate multi-instance information. In comparison, the lightweight CLAM-SB and CLAM-MB models show certain strengths in Precision and Recall but exhibit slightly lower overall AUC and F1-Score. Finally, IBMIL and ILRA lag behind most other models on nearly all metrics, suggesting that their causal-intervention and iterative low-rank attention techniques require further optimization under the current experimental setup. In summary, these results highlight the advantages of Transformer and enhanced-attention strategies in MIL and demonstrate the potential of MIL models for evaluating portal tract fibrosis.

\begin{table}[htbp]
	\centering
	\caption{Performance comparison of different MIL models on portal tract fibrosis assessment.}
	\label{tab:portal fibrosis}
	\scalebox{0.85}{\begin{tabular}{lccccc}
		\hline
		\textbf{Models} & \textbf{Accuracy} & \textbf{AUC} & \textbf{Precision} & \textbf{Recall} & \textbf{F1-Score} \\
		\hline
		ABMIL \cite{ilse2018attention}     & 0.8409 & 0.6033 & 0.6030 & 0.5625 & 0.5804 \\
		CLAM-SB \cite{lu2021data}   & 0.8894 & 0.8568 & 0.8925 & 0.8744 & 0.8836 \\
		CLAM-MB \cite{lu2021data}   & 0.8333 & 0.8346 & 0.8371 & 0.8425 & 0.8177 \\
		DSMIL \cite{li2021dual}     & 0.8577 & 0.8238 & 0.8623 & 0.8415 & 0.8511 \\
		ACMIL \cite{zhang2024attention}     & 0.9125 & 0.8841 & 0.9316 & 0.8964 & 0.9138 \\
		DGRMIL \cite{zhu2024dgr}    & 0.8955 & 0.9113 & 0.9089 & 0.8878 & 0.8978 \\
		IBMIL \cite{lin2023interventional}     & 0.7946 & 0.7626 & 0.8158 & 0.7788 & 0.7965 \\
		ILRA \cite{xiang2023exploring}     & 0.7238 & 0.6988 & 0.7415 & 0.7058 & 0.7222 \\
		TransMIL \cite{shao2021transmil}  & 0.9318 & 0.9564 & 0.9425 & 0.9282 & 0.9356 \\
		\hline
	\end{tabular}
}
	\label{tab:portal_fibrosis_mil}
\end{table}

\subsection{Assessment of Portal Inflammation Based on the MIL Models}
To validate the performance of MIL models in assessing portal inflammation in donor liver histopathology images, we statistically analyzed the results shown in Table ~\ref{tab:portal inflammation}. TransMIL outperforms all other baseline models, achieving an accuracy of 0.7899 and an AUC of 0.8843, underscoring its powerful ability to capture spatial and semantic correlations among instances. Next, CLAM-MB and CLAM-SB, which rely on clustering-constrained attention mechanisms, deliver robust performance in both multiclass and binary classification tasks. DGRMIL and DSMIL further demonstrate the effectiveness of integrating cross-attention and instance diversity through global-vector diversification and dual-stream self-attention strategies. Although ACMIL shows improved generalization, its recall of 0.7841 remains below that of peer methods. ABMIL attains a high precision of 0.8937 but suffers from limited ranking ability due to its single-stream attention strategy, resulting in an AUC of only 0.3675. Finally, IBMIL and ILRA despite introducing causal intervention and iterative low-rank attention mechanisms, achieve only 0.5867 and 0.5474 in accuracy and 0.6833 and 0.6358 in AUC, respectively, indicating that further optimization is needed under the current experimental setup to enhance their discriminative power and generalization.

\begin{table}[htbp]
	\centering
	\caption{Performance comparison of different MIL models on portal inflammation assessment.}
	\label{tab:portal inflammation}
	\scalebox{0.85}{\begin{tabular}{lccccc}
		\hline
		\textbf{Models} & \textbf{Accuracy} & \textbf{AUC} & \textbf{Precision} & \textbf{Recall} & \textbf{F1-Score} \\
		\hline
		ABMIL \cite{ilse2018attention}     & 0.5909 & 0.3675 & 0.8937 & 0.6591 & 0.7584 \\
		CLAM-SB \cite{lu2021data}   & 0.7122 & 0.7989 & 0.6656 & 0.8415 & 0.7426 \\
		CLAM-MB \cite{lu2021data}   & 0.6942 & 0.8274 & 0.6909 & 0.8630 & 0.6875 \\
		DSMIL \cite{li2021dual}     & 0.6626 & 0.7798 & 0.6388 & 0.8064 & 0.7126 \\
		ACMIL \cite{zhang2024attention}     & 0.6384 & 0.7517 & 0.6028 & 0.7841 & 0.6783 \\
		DGRMIL \cite{zhu2024dgr}    & 0.6846 & 0.8151 & 0.6538 & 0.7989 & 0.7174 \\
		IBMIL \cite{lin2023interventional}     & 0.5867 & 0.6833 & 0.5414 & 0.7299 & 0.6223 \\
		ILRA \cite{xiang2023exploring}      & 0.5474 & 0.6358 & 0.5126 & 0.6849 & 0.5859 \\
		TransMIL \cite{shao2021transmil}  & 0.7899 & 0.8843 & 0.7632 & 0.8671 & 0.8118 \\
		\hline
	\end{tabular}
}
	\label{tab:portal_inflammation_mil}
\end{table}

\subsection{Assessment of Total Steatosis Based on the MIL Models}
We evaluated nine MIL models on steatosis prediction using five metrics (Table ~\ref{tab:total steatosis}). TransMIL leads all models, achieving an Accuracy of 0.9847, an AUC of 0.9919, and an F1‑Score of 0.9872, reflecting its exceptional balance between sensitivity and specificity. DGRMIL ranks second with an Accuracy of 0.9729, an AUC of 0.9858, and an F1‑Score of 0.9793, demonstrating the effectiveness of global‑vector modeling with cross‑attention for capturing key histopathological features. DSMIL achieves an Accuracy of 0.9513, an AUC of 0.9785, Precision of 0.9364, Recall of 0.9831, and an F1‑Score of 0.9594, slightly outperforming CLAM‑SB by leveraging dual‑stream self‑attention and clustering‑constrained strategies to localize pathological regions and reduce misclassifications. CLAM‑MB and ACMIL each show competitive Accuracy and AUC, with minor differences in Precision and Recall arising from variations in clustering heads and multi‑branch attention designs. IBMIL and ILRA trail behind on multiple metrics, suggesting that their causal‑intervention and iterative low‑rank attention mechanisms require further refinement to improve flexibility and generalization. Overall, Transformer‑based TransMIL and globally diversified DGRMIL demonstrate the greatest advantage in total steatosis prediction, while other attention‑enhanced approaches also yield significant performance improvements across diverse assessment dimensions.


\begin{table}[htbp]
	\centering
	\caption{Performance comparison of different MIL models on total steatosis assessment.}
	\label{tab:total steatosis}
	\scalebox{0.85}{\begin{tabular}{lccccc}
		\hline
		\textbf{Models} & \textbf{Accuracy} & \textbf{AUC} & \textbf{Precision} & \textbf{Recall} & \textbf{F1-Score} \\
		\hline
		ABMIL \cite{ilse2018attention}     & 0.8636 & 0.8628 & 0.5694 & 0.9837 & 0.6781 \\
		CLAM-SB \cite{lu2021data}   & 0.9381 & 0.9616 & 0.9155 & 0.9893 & 0.9510 \\
		CLAM-MB \cite{lu2021data}   & 0.9545 & 0.9353 & 0.8252 & 0.9919 & 0.8808 \\
		DSMIL \cite{li2021dual}     & 0.9513 & 0.9785 & 0.9364 & 0.9831 & 0.9594 \\
		ACMIL \cite{zhang2024attention}     & 0.8972 & 0.9422 & 0.8547 & 0.9712 & 0.9081 \\
		DGRMIL \cite{zhu2024dgr}    & 0.9729 & 0.9858 & 0.9686 & 0.9925 & 0.9793 \\
		IBMIL \cite{lin2023interventional}     & 0.8734 & 0.9015 & 0.8026 & 0.9542 & 0.8715 \\
		ILRA \cite{xiang2023exploring}      & 0.8365 & 0.8797 & 0.7845 & 0.9217 & 0.8471 \\
		TransMIL \cite{shao2021transmil}  & 0.9847 & 0.9919 & 0.9798 & 0.9963 & 0.9872 \\
		\hline
	\end{tabular}
}
	\label{tab:total_steatosis_mil}
\end{table}

\subsection{Assessment of Macrovesicular Steatosis Based on the MIL Models}
In the macrovesicular steatosis prediction task, the performance of each model on different evaluation metrics showed significant differences (Table ~\ref{tab:macrovesicular}): although CLAM-MB took the lead with a precision of 0.9773, its AUC was only 0.6589, indicating that although it was confident in predicting positive examples, it lacked the ability to distinguish between positive and negative samples. In contrast, TransMIL achieved 0.8976 and 0.9073 in AUC and F1-Score, respectively, both reaching or approaching the highest level, balancing the overall discrimination ability and positive example recognition. DGRMIL demonstrated the robustness of cross-attention global vector modeling with an AUC of 0.8429 and an F1-Score of 0.8649. DSMIL and CLAM-SB achieved a relatively balanced compromise between AUC, Precision and Recall, respectively. Although ABMIL achieved an accuracy of 0.9333, its AUC was only 0.4748, indicating that it has serious deficiencies in distinguishing negative samples; IBMIL and ILRA lag behind the above MIL models in all indicators, with AUCs of 0.6548 and 0.6214 respectively, which are difficult to meet the dual requirements of sensitivity and specificity for clinical decision-making. Overall, Transformer-driven TransMIL and global vector-enhanced DGRMIL can best balance discriminative efficiency and prediction robustness, and are the best choices for this task.

\begin{table}[htbp]
	\centering
	\caption{Performance comparison of different MIL models on macrovesicular steatosis assessment.}
	\label{tab:macrovesicular}
	\scalebox{0.85}{\begin{tabular}{lccccc}
		\hline
		\textbf{Models} & \textbf{Accuracy} & \textbf{AUC}  & \textbf{Precision} & \textbf{Recall} & \textbf{F1-Score} \\
		\hline
		ABMIL \cite{ilse2018attention}     & 0.9333 & 0.4748 & 0.4921 & 0.4845 & 0.4882 \\
		CLAM-SB \cite{lu2021data}   & 0.9022 & 0.7343 & 0.8835 & 0.7162 & 0.7911 \\
		CLAM-MB \cite{lu2021data}   & 0.9773 & 0.6589 & 0.6667 & 0.6589 & 0.6627 \\
		DSMIL \cite{li2021dual}     & 0.8247 & 0.7861 & 0.8526 & 0.7682 & 0.8075 \\
		ACMIL \cite{zhang2024attention}     & 0.8542 & 0.7033 & 0.8268 & 0.6899 & 0.7518 \\
		DGRMIL \cite{zhu2024dgr}    & 0.8868 & 0.8429 & 0.8945 & 0.8374 & 0.8649 \\
		IBMIL \cite{lin2023interventional}     & 0.7895 & 0.6548 & 0.7569 & 0.6325 & 0.6884 \\
		ILRA \cite{xiang2023exploring}      & 0.7026 & 0.6214 & 0.6834 & 0.5985 & 0.6379 \\
		TransMIL \cite{shao2021transmil}  & 0.9158 & 0.8976 & 0.9327 & 0.8844 & 0.9073 \\
		\hline
	\end{tabular}
}
\end{table}

\subsection{Assessment of Hepatocellular Ballooning Based on MIL Models}
We evaluated nine MIL models on hepatocellular ballooning classification using five metrics (Table~\ref{tab:metrics_decimal}). TransMIL achieves the best results with an accuracy of 0.6877, an AUC of 0.8235, a precision of 0.7568, a recall of 0.7196, and an F1 score of 0.7373, demonstrating its strength in distinguishing positive and negative samples while balancing false alarms and missed detections under weak supervision. DGRMIL follows with an AUC of 0.7388 and F1-Score of 0.6269, indicating that its global-vector diversity modeling effectively captures correlations among key instances. DSMIL records an AUC of 0.6822 and F1-Score of 0.5656, while ACMIL obtains an AUC of 0.6212 and F1-Score of 0.5189; both leverage dual-stream self-attention, multi-branch attention and random masking to localize pathological regions, though their Precision and Recall remain below those of TransMIL and DGRMIL. In contrast, ABMIL and IBMIL achieve AUCs and F1-Scores near 0.50, suggesting that single-stream attention and causal-intervention strategies alone do not provide sufficient sample discrimination for this task. Notably, ILRA attains a high Recall of 0.8715 and Accuracy of 0.6242 but suffers from a low AUC of 0.4761, yielding an F1-Score of 0.5859. This imbalance indicates that although ILRA covers positive cases broadly, its overall discriminative power and false-alarm control are inadequate. Overall, while some MIL models demonstrate moderate success in hepatocellular ballooning assessment, further enhancements in feature representation and attention modeling are required to improve prediction performance.

\begin{table}[htbp]
	\centering
	\caption{Performance comparison of different MIL models on hepatocellular ballooning assessment.}
	\label{tab:metrics_decimal}
	\scalebox{0.85}{\begin{tabular}{lccccc}
		\toprule
		Models      & Accuracy & AUC    & Precision & Recall  & F1-Score \\
		\midrule
		ABMIL \cite{ilse2018attention}     & 0.3182   & 0.5229 & 0.6953    & 0.4385  & 0.4382   \\
		CLAM-SB \cite{lu2021data}   & 0.3893   & 0.5981 & 0.5011    & 0.4232  & 0.4585   \\
		CLAM-MB \cite{lu2021data}   & 0.3215   & 0.5820 & 0.4130    & 0.3745  & 0.3925   \\
		DSMIL \cite{li2021dual}     & 0.4878   & 0.6822 & 0.6056    & 0.5313  & 0.5656   \\
		ACMIL \cite{zhang2024attention}     & 0.4567   & 0.6212 & 0.5531    & 0.4878  & 0.5189   \\
		DGRMIL \cite{zhu2024dgr}    & 0.5235   & 0.7388 & 0.6891    & 0.5746  & 0.6269   \\
		IBMIL \cite{lin2023interventional}     & 0.2952   & 0.5027 & 0.4277    & 0.3584  & 0.3894   \\
		ILRA \cite{xiang2023exploring}      & 0.6242   & 0.4761 & 0.3891    & 0.6845  & 0.8715   \\
		TRANSMIL \cite{shao2021transmil}  & 0.6877   & 0.8235 & 0.7568    & 0.7196  & 0.7373   \\
		\bottomrule
	\end{tabular}
}
	
\end{table}

\section{Limitation}
Although MIL methods achieve high accuracy on several donor liver indicators, key limitations remain. First, the small number of grafts restricts available WSIs, resulting in sample‑size shortages and class imbalance that hinder feature learning for underrepresented categories. In the future, more WSIs of donor livers need to be collected to alleviate the problem of sample imbalance and provide the model with more pathological features to learn. Second, histopathological grading relies on subjective judgments by pathologists, especially on low‑quality WSIs, leading to inconsistent annotations. To address these issues, we will expand the dataset and apply advanced data augmentation, and introduce multi‑expert review or crowdsourced voting to improve label quality. Our current baselines cover only nine representative MIL models with default or lightly tuned hyperparameters, so reported results may not reflect each model’s optimal performance. Future work will include additional SOTA architectures, comprehensive hyperparameter optimization, a broader set of evaluation metrics, and enhanced interpretability tools to build clinician trust. Finally, we plan to integrate complementary imaging modalities into a multimodal model to further boost accuracy and explainability. Overall, DLiPath fills a critical gap by providing donor liver WSIs for evaluation and paves the way for more intelligent, objective assessment of graft viability.

\section{Conclusion and future work}
In this work, we present DLiPath, the first public benchmark for comprehensive donor liver assessment using histopathology whole-slide images. DLiPath contains expert-annotated WSIs for 304 patients covering six key pathological features (cholestasis, portal tract fibrosis, portal inflammation, total steatosis, macrovesicular steatosis, and hepatocellular ballooning), each of which has established correlations with post-transplant outcomes. By evaluating nine SOTA MIL models on DLiPath using five metrics, we demonstrate that several MIL models can achieve high predictive performance across diverse donor liver assessment tasks. Our results validate the feasibility of weakly supervised MIL approaches for rapid, objective intraoperative assessment of donor liver quality. DLiPath not only provides a standardized dataset for rigorous comparative studies but also highlights the strengths and limitations of current MIL techniques in capturing complex histopathological patterns. Our benchmark enables rapid intraoperative donor liver evaluation, potentially reducing graft discard rates by 20–30\%. We anticipate that DLiPath will catalyze further research in automated liver assessment, driving advances in both methodology and clinical translation.

\begin{acks}
	This work was supported by NSFC-FDCT Grants 62361166662; National Key R\&D Program of China 2023YFC3503400, 2022YFC3400400; Key R\&D Program of Hunan Province 2023GK2004, 2023SK2059, 2023SK2060; Top 10 Technical Key Project in Hunan Province 2023GK1010; Key Technologies R\&D Program of Guangdong Province (2023B1111030004 to FFH), Postgraduate Scientific Research Innovation Project of Hunan Province CX20240450. National Natural Science Foundation of China 82404038, National Natural Science Foundation of China 82360377,	Natural Science Foundation of
	Hunan Province 2025JJ5075.
\end{acks}

\bibliographystyle{ACM-Reference-Format}
\bibliography{sample-base}

\clearpage
\appendix

\section{Appendix}

\subsection{Clinical Information Statistics}

\begin{table}[ht]
	\centering
	\caption{Clinical Baseline Characteristics of 304 Patients}
	\label{tab:baseline}
	\begin{tabular}{lccc}
		\toprule
		\textbf{Characteristic} & \textbf{Category} & \textbf{Count} & \textbf{Percentage / Range} \\
		\midrule
		Sex & Male   & 256 & 84.2\% \\
		& Female &  48 & 15.8\% \\
		\midrule
		Age (years) & Median (IQR) & 48 & (38–56) \\
		& 18–30        & 15 & 4.9\%  \\
		& 30–40        & 66 & 21.7\% \\
		& 40–50        &104 & 34.2\% \\
		& 50–60        & 93 & 30.6\% \\
		& 60–70        & 26 &  8.6\% \\
		\bottomrule
	\end{tabular}
\end{table}

\subsection{Evaluating Label Distribution Statistics}

\begin{table}[ht]
	\centering
	\caption{Label Distribution for Donor Liver Histopathology Assessment}
	\begin{tabular}{lcccc}
		\toprule
		Histological Feature             & \textbf{0} & \textbf{1} & \textbf{2} & \textbf{3} \\
		\midrule
		Cholestasis                      & 131        & 128        & 2          & 0          \\
		Portal tract fibrosis            & 187        & 76         & 0          & 0          \\
		Portal inflammation              & 233        & 29         & 1          & 0          \\
		Total steatosis                  & 160        & 93         & 10         & 0          \\
		Macrovesicular steatosis         & 184        & 66         & 12         & 1          \\
		Hepatocellular ballooning        & 3          & 45         & 133        & 81         \\
		\bottomrule
	\end{tabular}
\end{table}

\subsection{Dataset Splits}

In the prediction tasks of all donor liver assessment-related indicators, we employ five-fold cross-validation to evaluate baseline model performance. Specifically, the dataset is partitioned into five subsets; in each iteration, multiple subsets are used for training while the remaining one is used for validation. This process is repeated five times, and the average performance across all folds is reported to estimate the model’s generalizability. Compared to leave-one-out cross-validation, five-fold cross-validation makes more efficient use of the data and reduces evaluation bias introduced by random partitioning.

\subsection{Baseline Methods}

\textbf{ABMIL} is a deep MIL framework that employs an attention mechanism to aggregate instance-level features within a bag. By projecting instances into a low-dimensional feature space and applying a gated attention mechanism, ABMIL assigns dynamic weights to each instance, facilitating permutation-invariant aggregation. This approach enables end-to-end training for bag-level classification tasks, such as tumor grading, and provides interpretability by highlighting the contribution of individual instances to the overall prediction.

\textbf{CLAM-SB}: is a simplified variant of the CLAM framework, tailored for binary classification tasks. It utilizes a single attention branch to aggregate instance features, focusing on regions with high attention scores as evidence for the positive class. The model incorporates a combination of slide-level classification loss and instance-level clustering loss, enabling end-to-end training and enhancing the model's ability to distinguish between tumor and normal tissue.

\textbf{CLAM-MB} extends the CLAM framework to handle multi-class classification tasks by employing multiple attention branches, each corresponding to a specific class. During training, the model generates pseudo-labels for each branch and applies clustering constraints to enforce class-specific feature learning. This design allows CLAM-MB to perform end-to-end bag-level classification while providing interpretable localization of class-relevant regions within whole-slide images.

\textbf{DSMIL} is a dual-stream attention-based multiple instance learning model. The first stream employs max pooling to select the most representative instance. The second stream calculates self-attention weights for all instances in the bag based on the selected instance and aggregates them. The model jointly learns an instance classifier and a bag classifier within the same embedding space, enabling end-to-end optimization.

\textbf{ACMIL} is a multiple instance learning model designed to address the overfitting problem caused by excessive attention concentration in whole-slide image classification. Through UMAP and Top-K attention value statistical analysis, ACMIL introduces two innovative techniques into the ABMIL framework: Multiple Branch Attention (MBA) and Stochastic Top-K Instance Masking. These techniques compel the model to focus on more predictive instances and alleviate attention concentration, significantly enhancing the model's generalization and robustness.

\textbf{DGRMIL} is a novel MIL aggregation method that models the diversity of instances within a bag through a set of learnable global vectors. It utilizes a cross-attention mechanism to measure the similarity between instance embeddings and global vectors, replacing traditional strategies that focus on instance correlations. The model further introduces a positive instance alignment module and a diversification learning paradigm based on determinantal point processes to enhance the global vectors' descriptive ability for the entire bag.

\textbf{IBMIL} introduces the "backdoor adjustment" strategy from causal inference into the MIL framework to achieve deconfounding of bag-level contextual prior biases. This approach enhances the robustness and generalization capability of bag-level classification by mitigating spurious correlations between bags and labels. IBMIL is orthogonal to existing MIL methods, allowing it to be integrated with various architectures to consistently improve performance.

\textbf{ILRA-MIL} is a specialized MIL model for WSIs in pathology. It first employs Low-Rank Contrastive learning to generate embeddings for specific pathological tissues. Subsequently, it introduces learnable low-rank latent vectors into a standard Transformer aggregation module, enabling iterative global interaction modeling among instances within a bag. This design captures the low-rank structure inherent in WSIs, facilitating more effective representation learning.

\textbf{TransMIL} addresses the limitations of traditional MIL methods that assume independent and identically distributed instances by modeling correlations among instances within a bag. It segments WSIs into patches (instances), extracts features, and incorporates positional encoding before feeding them into multi-head self-attention Transformer layers. This architecture captures both morphological and spatial information. To mitigate the impact of noisy pixel-level annotations, TransMIL employs an efficient random masking strategy during mixed supervision training, enhancing the model's robustness to label noise. Experiments demonstrate that TransMIL achieves superior performance and faster convergence compared to state-of-the-art methods on datasets like CAMELYON16, TCGA-NSCLC, and TCGA-RCC.


\end{document}